\documentclass[]{pasj01}

\begin{document} 
\Received{2018/05/18}
\Accepted{}

\title{Is the Infrared Background Excess Explained by the Isotropic Zodiacal Light from the Outer Solar System?}

\author{Kohji \textsc{Tsumura}\altaffilmark{1}%
\altaffiltext{1}{Frontier Research Institute for Interdisciplinary Science, Tohoku University, \\ 6-3 Aramaki Aza-Aoba, Aoba-ku, Sendai, 980-8578, JAPAN}
\email{tsumura@astr.tohoku.ac.jp}}

\KeyWords{interplanetary medium, zodiacal dust, infrared: diffuse background} 

\maketitle

\begin{abstract}
This paper investigates whether an isotropic zodiacal light from the outer Solar system can account for the detected background excess in near-infrared.
Assuming that interplanetary dust particles are distributed in a thin spherical shell at the outer Solar system ($>200$ AU), 
thermal emission from such cold ($<30$ K) dust in the shell has a peak at far-infrared ($\sim 100$ \micron).
By comparing the calculated thermal emission from the dust shell with the observed background emissions at far-infrared, permissible dust amount in the outer Solar system is obtained.
Even if the maximum dust amount is assumed, the isotropic zodiacal light as the reflected sunlight from the dust shell at the outer Solar system cannot explain the detected background excess at near-infrared.
\end{abstract}

\section{Introduction}
The Extragalactic Background Light (EBL) arises from integrated emission from the first star production era to the present day.
Thus observation of EBL as the accumulated history of the universe is important to understand the star formation history.
Recent observations show that the measured EBL at optical and near-infrared (NIR) has an excess of $\sim 30$ nW/m$^2$/sr over the cumulative light from galaxies 
\citep{Tsumura, Matsumoto, Sano1, Sano2, Mattila, CIBER},
meaning that there are unknown light sources in the universe.
For candidates of light sources for this NIR background excess, intra-halo light \citep{Cooray, Zemcov}, emissions from LIGO-type blackholes \citep{Kashlinsky},
and decaying hypothetical particles \citep{Kohri} are proposed.

On the other hand, isolation of the EBL from foreground emissions is difficult due to its diffuse, extended nature. 
The largest uncertainty comes from the removal of the dominant foreground, the zodiacal light, 
which is scattered sunlight at optical and NIR, and thermal emission at mid- and far-infrared (FIR) from interplanetary dust (IPD) within the Solar system.
In this reason, some authors claimed that the NIR background excess is caused by systematic uncertainty in subtraction of the zodiacal light \citep{Dwek, Kawara}. 
In the recent EBL observations, the zodiacal light is subtracted using the model based on morphology and time variation measured by DIRBE on {\it COBE} satellite \citep{Kelsall}.
If there is an isotropic zodiacal light component showing no time variation, such isotropic component is not included in the zodiacal light model.
Thus the isotropic zodiacal light component can be a source of the NIR background excess \citep{Dwek2, Chary}.  
   
Such isotropic zodiacal light component, if it exists, is made by IPD around the Earth or around the outer Solar system, because it does not show time variation by the observations from the Earth.
\citet{Matsumoto2} compared the zodiacal light model \citep{Kelsall} with the observational data of the zodiacal light during the cruising to Jupiter by {\it Pioneer 10},
and no evidence was found that the isotropic zodiacal light component exists around the Earth. 
In this reason, the other possibility of the existence of IPD around the outer Solar system is investigated in this paper.

The zodiacal light is dominated by the thermal emission from the nearby IPD whose temperature is $\sim$280 K,
and the peak of such thermal emission comes at mid-infrared.
On the other hand, if the dust exists around the outer Solar system to make the isotropic zodiacal light component, 
temperature of such IPD is quite low, and thermal emission from such low temperature IPD has a peak at FIR.
Thus, as the strategy of this paper, it is investigated whether the NIR background excess is explained by the isotropic zodiacal light from the IPD around the outer Solar system,
whose amount is restricted by the FIR background data.

This paper is organized as follows. 
In Section 2, the acceptable upper limit of IPD amount is obtained from the FIR background data.
Using this amount of IPD, isotropic zodiacal light component at NIR is calculated, and it is compared with the NIR background excess in Section 3.
Then discussion on our results comes in Section 4, and conclusion comes in Section 5.

\section{Far Infrared Background Excess}
\subsection{Dust Emission Formulation}
Here the method to calculate the thermal emission from IPD in the outer Solar system used in this paper is explained.

For simplicity, it is assumed that IPD are distributed as a thin spherical shell at distance $d$ from the Sun.
This assumption is supported by the fact that most of long period comets with semi-major axes of $>40$ AU have isotropic inclination.
When the total number of the IPD particles in the spherical shell is $N$, its surface number density is $n = N/4\pi$ sr$^{-1}$.
Thermal emission flux from an IPD particle with radius $a$ at distance $d$ can be written as $(a/d)^2 \epsilon_{\lambda} \pi B_{\lambda}(T_{dust})$,
where $\epsilon_{\lambda}$ is the emissivity of the IPD 
and $B_{\lambda}(T_{dust})$ is the Planck function in the unit of W/m$^2$/\micron\ at wavelength $\lambda$ and IPD temperature $T_{dust}$.
The usual dust emissivity law is
\begin{equation}
  \epsilon_{\lambda} = \left( \frac{2 \pi a}{\lambda} \right) ^{\beta}
\end{equation}
where the power-low index $\beta$ has values between 1 and 2 , and $\beta =1$ is the case for amorphous dusts and $\beta =2$ is for metal and crystal dusts.
We adopt $\beta =2$ from the zodiacal light observation by DIRBE/{\it COBE} \citep{Fixsen}.
By combining these, the thermal emission from the spherical IPD shell, $\lambda I_{\lambda}^{FIR}$ W/m$^2$/sr, can be written as 
\begin{equation}
  \label{isoZL_eq}
  \lambda I_{\lambda}^{FIR} = \frac{\pi n \epsilon_{\lambda} a^2}{d^2} \lambda B_{\lambda}(T_{dust}) = \frac{4 \pi^3 n a^4}{d^2} \frac{B_{\lambda}(T_{dust})}{\lambda}
\end{equation}

\begin{figure}
 \begin{center}
  \includegraphics[width=8cm]{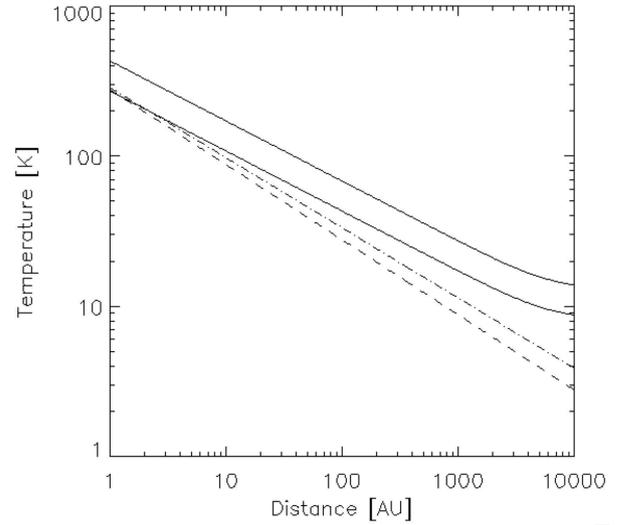} 
 \end{center}
\caption{Radial profile of the dust temperature. The solid lines show the Stern case (equation \ref{Stern_eq}) of a=10 \micron\ (lower) and 1 \micron\ (upper) cases \citep{Stern}, 
              the dashed line shows the Mann case (equation \ref{Mann_eq}) \citep{Mann}, and the dash-dot line shows the Kelsall case \citep{Kelsall}.}
\label{temp}
\end{figure}

By assuming thermal equilibrium, the dust temperature $T_{dust}$ can be written as
\begin{equation}
 \label{Mann_eq}
  T_{dust}(d) = T_{\odot} \left( \frac{1-A}{4} \right)^{1/4} \left( \frac{R_{\odot}}{d} \right)^{1/2}
\end{equation}
where $A$ is the typical albedo of IPD, and $T_{\odot}$ and $R_{\odot}$ is the temperature and radius of the Sun \citep{Mann}. 
Adopting $A=0.05$ as a typical value of cometary dust \citep{Hanner}, this equation (\ref{Mann_eq}) gives $T_{dust} = 276 [K] \cdot d[AU] ^{-0.5}$ , 
which is consistent with the result of $T_{dust}(d)= 286 [K] \cdot d [AU] ^{-0.467}$ from the zodiacal light observation with DIRBE/{\it COBE} \citep{Kelsall}. 
At the outer solar system, however, the ambient interstellar starlight and the cosmic microwave background (CMB) are not negligible in the thermal contribution 
relative to the insolation from the Sun.
\citet{Stern} gave an equation to express the IPD temperature in such a situation as below
\begin{equation}
 \label{Stern_eq}
  T_{dust}(d) = 5.2 \left[  (1-A) \left( \frac{1.5a}{1\ [\micron]}\right)^{-1} \left( T_{bg}^4 + \frac{L_{\odot}}{16\pi\sigma_{SB}d^2}\right) \right]^{1/5}
\end{equation}
where $\sigma_{SB}$ is the Stefan-Boltzmann constant, $L_{\odot}$ is the Solar luminosity,
and $T_{bg}$ is the equivalent blackbody temperature of the background flux from ambient starlight and CMB, adopting $T_{bg} = 3.5$ K \citep{Spitzer}.
Figure \ref{temp} compares the radial profiles of the IPD temperature $T_{dust}$ of each case.
Stern case (equation \ref{Stern_eq}) of $a=10$ \micron\ agrees well with the Mann case (equation \ref{Mann_eq}) and the Kelsall case at around 1 AU,
which confirms that the dominant IPD size is 10-100 \micron\ from the zodiacal light observations \citep{Grun}. 
Hereafter, the Stern case (equation \ref{Stern_eq} and solid lines in Figure \ref{temp}) is used as the IPD temperature in this paper.

\begin{figure*}
 \begin{center}
  \includegraphics[width=16cm]{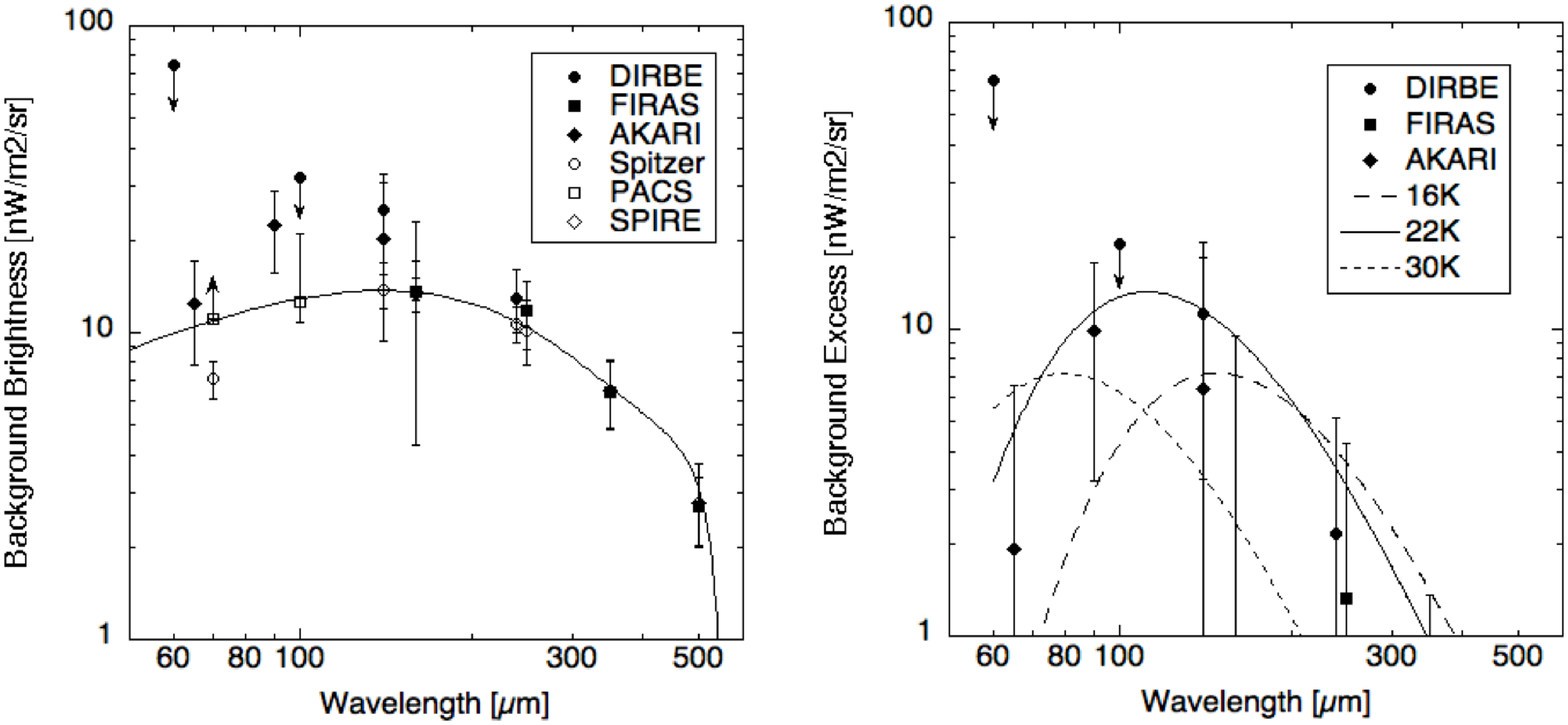} 
 \end{center}
\caption{Left: FIR background brightness. Filled symbols are values of direct measurements: 
filled circles are DIRBE/{\it COBE} \citep{Odegard}, filled squares are FIRAS/{\it COBE} \citep{Lagache}, and filled diamonds are FIS/{\it AKARI} \citep{Matsuura}. 
Open symbols are values of integrated light of galaxies: 
open circles are MIPS/{\it Spitzer} \citep{Dole, Odegard}, open squares are PACS/{\it Herschel} \citep{Berta}, and open diamonds are SPIRE/{\it Hearschel} \citep{Bethermin}.
The solid line shows a polynomial fitting of the integrated light of galaxies (open symbols).
Right: FIR background excess of the directly measured values over the integrated light of galaxies.
Thermal emissions from the spherical IPD shell at the outer Solar system described in equation (\ref{isoZL_eq}) are also shown in 
dashed curve (16 K, Case A), solid curves (22 K, Case B), and dotted curve (30 K, Case C).}
\label{FIRCIB}
\end{figure*}

By assuming all IPD particles have the same size $a$, temperature of the IPD $T_{dust}$ is uniquely defined from the distance to the spherical IPD shell $d$ from equation (\ref{Stern_eq}) .
Thus the spectral shape of the thermal emission is defined by $T_{dust}$ (or $d$), and it is scaled by the IPD amount $n$ (see equation (\ref{isoZL_eq})). 
The total IPD mass in the spherical shell $M_{shell}$ can be written as $M_{shell} = 4\pi a^3 \rho N /3 = 16\pi^2 a^3 \rho n /3$, 
where $\rho$ is the average IPD mass density and $\rho = 2$ g/cm$^3$ is adopted for a typical mass density of IPD.

\begin{table*}
  \tbl{Fitting results of the thermal emissions from the IPD shell to the FIR background excess.}{%
  \begin{tabular}{cccccc}
  \hline\noalign{\vskip3pt} 
  Case & IPD size $a$ [\micron] & IPD Temperature $T_{duat}$ [K] & Distance to the IPD shell $d$ [AU] $^*$ & Total IPD mass $M_{shell} / M_{Earth}$$^{**}$ & $\lambda I_{\lambda}^{NIR}$ at 1 \micron\ [nW/m$^2$/sr]\\  
   \hline\noalign{\vskip3pt} 
  A & 10 & 16 & 1150 & $1.1 \times 10^{-1}$  & 0.028\\
  B & 10 & 22 & 560 & $8.5 \times 10^{-3}$  & 0.040\\
  C & 10 & 30 & 240 & $1.1 \times 10^{-4}$ & 0.016\\
  A' & 1  & 16 & 4340 & $1.5 \times 10^1$   & 0.202\\
  B' & 1  & 22 & 1840 & $9.1 \times 10^{-1}$ & 0.378\\
  C' & 1  & 30 & 760 & $1.1 \times 10^{-2}$ & 0.158\\
  \hline\noalign{\vskip3pt} 
  \end{tabular}}
  \begin{tabnote}
    \hangindent6pt\noindent
    \hbox to6pt{\footnotemark[$*$]\hss}\unskip%
     Distance to the dust shell $d$ is derived from the equation (\ref{Stern_eq}) by the dust temperature $T_{dust}$.\\
      \hangindent0pt \footnotemark[$**$]\hss\unskip%
      Dust mass density $\rho =2$ g/cm$^3$ is adapted.
\end{tabnote}
  \label{table}
\end{table*}

\subsection{Far-Infrared Data}
There are two methods to derive the FIR background brightness.
One is to sum up the flux of resolved galaxies and extrapolated fainter unresolved objects, 
which gives the lower estimate of FIR background brightness because only the identified light sources are into account.
Open symbols in Figure \ref{FIRCIB} (left) denote the values in this method by PACS \citep{Berta} and SPIRE \citep{Bethermin} on {\it Herschel} Space Telescope,
and MIPS on {\it Spitzer} Space Telescope \citep{Dole, Odegard}, and a solid curve shows a polynomial fitting of these values.
Number counts of galaxies obtained by {\it Herschel} were fitted by power-low functions,
and they were integrated down to 1 $\mu$Jy at $<160$ \micron\ by PACS \citep{Berta}, and down to zero flux at $>250$ \micron\ by SPIRE \citep{Bethermin}.
Although the obtained FIR background value at 70 \micron\ is a lower limit because the curve obtained from the power-law fit is not fully converging at 1 $\mu$Jy,
the other values by PACS are regarded as equivalent to integration down to zero flux because fitted curves are converging enough \citep{Berta}.
The FIR background values from {\it Spitzer} was obtained from stacked images of 24 \micron\ sources with $>$60 $\mu$Jy at 70 and 160 \micron\ \citep{Dole},
and it was scaled to 140 and 240 \micron\ using a spectral energy distribution model \citep{Odegard}.

The other method is the direct measurements of the sky brightness.
This method gives the upper estimate of FIR background brightness because of a big difficulty in subtracting the foreground emissions 
from our Solar system (zodiacal light) and our Galaxy (diffuse Galactic light).
Filled symbols in Figure \ref{FIRCIB} (left) denote the values in this method by DIRBE \citep{Odegard} and FIRAS \citep{Lagache} on {\it COBE} satellite,
and FIS on {\it AKARI} satellite \citep{Matsuura}.
First, the zodiacal light was removed from the directly observed sky brightness based on the zodiacal light model \citep{Kelsall}.
Then, Galactic foreground emissions were subtracted based on the correlation with the HI 21 cm line data \citep{Snowden, Elvis, Stark},
the H-$\alpha$ total intensity data \citep{Reynolds, Haffner}, and the 100 \micron\ dust thermal emission data \citep{Schlegel}.

As shown in Figure \ref{FIRCIB} (left), the values derived between these two methods are basically consistent with each others,
but there is a small excess of the FIR background brightness over the integrated light of galaxies at around 100 \micron.
Figure \ref{FIRCIB} (right) shows the FIR background excess derived by the difference between the directly measured values and the interoperated values of the integrated light of galaxies.
The zodiacal light model uncertainties to the FIR background are 26.7, 6.3, 2.3, and 0.5 nW/m$^2$/sr at 60, 100, 140, and 240 \micron, respectively \citep{Kelsall},
thus the obtained FIR background excess cannot be explained by the model uncertainty. 
The origin of this FIR background excess is still unknown, but a contribution of luminous dusty galaxies with hot dust ($\sim$ 60 K) at z = 2-3 is discussed \citep{Blain}.

\subsection{Upper Limit of IPD Amount at Outer Solar System}
This FIR background excess can be a room for the isotropic zodiacal light component.
Assuming that all of the FIR background excess is caused by the isotropic zodiacal light, 
an upper limit of the IPD amount in the outer Solar system was obtained
by fitting this FIR background excess with the thermal emission from the spherical IPD shell at the outer Solar system.

Table \ref{table} shows the obtained fitting parameters in the equation (\ref{isoZL_eq}), and fitted curves are shown in Figure \ref{FIRCIB} (right).
Obtained best temperature of IPD is $T_{dust} \sim $22 K to reproduce the peak wavelength of the FIR background excess at  $\sim100$ \micron,
corresponding to the distance to the spherical IPD shell of $d \sim 560$ AU in the dust size $a = 10$ \micron\ case from Figure \ref{temp}.
In this configuration, the upper limit of the total mass of the spherical IPD shell $M_{shell}/M_{Earth} < 8.5 \times 10^{-3}$ (Case B) is obtained, 
where $M_{Earth} = 5.97 \times 10^{24}$ kg is the mass of the Earth.

Although the IPD temperature of 22 K is the most likely, 
temperature range between 16 K and 30 K is allowed within the errors of FIR background excess as shown in Figure \ref{FIRCIB} (right).
When colder IPD temperature is adopted, required distance from the Sun $d$ become larger, 
thus more masses are allowed within the FIR background excess.
In this reason, more conservative upper limit of the total IPD mass $M_{shell}/M_{Earth} < 1.1 \times 10^{-1} $ is obtained (Case A).

\section{Near Infrared Background Excess}
Next, the isotropic zodiacal light at NIR is calculated as the reflected sunlight from the spherical IPD shell obtained above.
The Solar flux at distance $d$ can be written as $L_{\lambda, \odot}/(4\pi d^2)$ where $L_{\lambda, \odot}$ is the Solar luminosity at wavelength $\lambda$, 
and IPD there receives this flux with the cross section $\pi a^2$.
The IPD particle scatters this light to $4\pi$ sr with the reflectance $A$ (albedo), and we on the Earth observe the scattered sunlight from IPD in the shell as the isotropic zodiacal light.
The distance between the Sun and the IPD particle, $d$, can be regarded as the distance between the Earth and the IPD, because of $d \gg 1$ AU.
The IPD particles are distributed in the spherical shell with the surface number density of $n$ sr$^{-1}$.
Given these, the reflected sunlight at NIR, $\lambda I_{\lambda}^{NIR}$, from the spherical IPD shell can be written as
\begin{equation}
  \label{NIRZL_eq}
  \lambda I_{\lambda}^{NIR} = n\frac{\lambda L_{\lambda, \odot}}{4\pi d^2}\pi a^2 A \Phi (\theta) \frac{1}{4\pi d^2} = \frac{n a^2 A \Phi (\theta)}{16 \pi d^4}\lambda L_{\lambda, \odot}
\end{equation}
where $\Phi (\theta)$ is a phase function.
Because the shell is far away from observer, $\Phi (\theta)$ = 0.2 is adopted for backward scattering from \citet{Kelsall}.
Using the Standard Solar Flux at 1 AU \citep{Tobiska}, the obtained NIR isotropic emissions are 0.028, 0.040, and 0.016 nW/m$^2$/sr at 1 \micron\ for Case A, B, and C as shown in Table \ref{table}, respectively.
These values are $>2$ orders lower than the reported values of NIR background excess at $\sim 1$ \micron\ \citep{CIBER}.
As a conclusion, even assuming the maximum amount of dust permitted from the FIR background excess, 
the NIR background excess cannot be explained by the isotropic zodiacal light.

\citet{Chary} stated that if they assume the existence of high-albedo IPD (ice mantle grains) with $A > 0.8$ with $\sim 50$ K at $\sim 40$ AU,
 the NIR background excess can be explained by the isotropic zodiacal light.
 However, such high temperature IPD with $\sim 50$ K is not consistent with the obtained FIR background excess in this paper (see Figure \ref{FIRCIB} (right)).

\section{Discussion}
\subsection{Parameter Dependence of $M_{shell}$}
The obtained total mass of the spherical IPD shell $M_{shell}$ in Table \ref{table} depends on the average IPD mass density $\rho$, and $\rho = 2$ g/cm$^3$ is adopted in this paper.
Mass density range of IPD particles (5-15 \micron\ size) collected at the stratosphere is 0.3 - 6.0 g/cm$^3$, averaging 2.0 g/cm$^3$ \citep{Love}.
\citet{Grun} also state that majority of the IPD have $\rho$ = 2 - 3 g/cm$^3$, whereas 20-40 \% of the IPD has low density ($<1$ g/cm$^3$).
Such low density IPD are though to be cometary origin \citep{Joswiak, Wiegert}.
In the outer Solar system, contributions of cometary origin IPD may be increased, and thus the average mass density may be smaller than our adopted value of $\rho = 2$ g/cm$^3$.
In such a case of smaller mass density, the required total mass of the spherical IPD shell $M_{shell}$ becomes low, thus the upper limit obtained in this paper becomes more conservative.

The obtained total IPD mass $M_{shell}$ also depends on the IPD size $a$, because smaller IPD are hotter, thus farther distance $d$ from the Sun is required to be 16-30 K (Figure \ref{temp}), 
and then more masses is allowed within the FIR background excess.
Results in the cases of $a = 1$ \micron\ are also shown in Table \ref{table} for comparison.
These results mean that if the IPD in the outer Solar system is dominated by small particles ($a = 1$ \micron), 
about 100 times more IPD masses than the case of $a = 10$ \micron\  can be acceptable in the outer Solar system within the FIR background excess,
resulting that about 10 times more NIR background $\lambda I_{\lambda}^{NIR}$ is obtained.
This result is consistent with the result in \citet{Dwek2} of  $10^{-3} < M_{shell} / M_{Earth} < 10^{2}$ at $> 700$ AU for $\beta =2$ case.
On the other hand, however, the total IPD mass is dominated by large particles ($\sim$100 \micron) from the dust size distribution around the Earth \citep{Kral}.
Although this dust size distribution is valid around the Earth, it is difficult to believe that the small IPD particles ($a < 1$ \micron) is dominated in the outer Solar system.
Anyway, even in the $a = 1$ \micron\ case with $M_{shell} \sim 10 M_{Earth}$ at $\sim$ 4300 AU (Case A'), 
the obtained isotropic zodiacal light at NIR $\lambda I_{\lambda}^{NIR}$ is still $< 1/10$ of the NIR background excess.
Therefore, the NIR background excess cannot be explained even in the small dust size cases.

 \subsection{Comparison to Models and Observations}
The maximum permissible IPD mass in the outer Solar system obtained in this study is much more than the total IPD mass inside Jupiter's orbit ($10^{-9}$ - $10^{-8}M_{Earth}$) \citep{Fixsen, Nesvorny}.
Is it realistic that such amount of IPD exist in the outer Solar system?
Outside of the heliosphere ($> 250$ AU), IPD are charged in the interstellar environment and ejected by the interstellar magnetic field.
According to \citet{Belyaev}, IPD with $a > 15$ \micron\ at $d > 1000$ AU and $a > 1$ \micron\ at $d > 100$ AU are ejected from the Solar system.
\citet{Dwek2} stated that only a cloud consisting of IPD larger than $\sim 1$ cm located between $\sim 5$ and 150 AU would be stable.
Therefore, the IPD shell assumed in this study cannot survive unless there is continuous supply of IPD.
IPD is believed to be supplied from comets \citep{Nesvorny, Yang} and asteroids \citep{Dermott, Nesvorny2, Tsumura2}, 
but comets are not active and asteroidal collisions are also less likely to occur in the cold and low density environment in the outer Solar system.
Therefore, the large amount of dust assumed in this study cannot be supplied by any of known mechanisms.

\citet{Poppe} constructed a dust density model based on the in-situ dust counting by {\it Pioneer 10}, {\it Galileo}, and {\it New Horizons}.
According to this model, dust density of 20 \micron\ size is $\sim 5 \times 10^{-4}$ km$^{-3}$ at 70 AU, 
dominated by dust grains originated from the Edgeworth-Kuiper belt objects and the Oort cloud comets.
Even by assuming this dust density is kept up to 1000 AU, the mass of a dust shell at 1000 AU with 10 AU thickness is $<3 \times 10^{-7} M_{Earth}$,
which is much less than the required amount to explain the FIR background excess (see Table \ref{table}).
   
In addition, a Solar-type star HD72905 (G1.5V, age $\sim 0.4$ Gyr) has 70 \micron\ excess ($L_{dust}/L_{\star} \sim 10^{-5}$) in its spectrum detected by MIPS/{\it Spitzer},
and this extra emission is produced by cool ($< 100$ K) dust of $<10^{-2} M_{Earth}$ \citep{Bryden}.
Because our Solar system has $L_{dust}/L_{\odot} \sim 2 \times 10^{-7}$ \citep{Nesvorny}, the IPD amount in our Solar system should be less than that in HD72905.

In these reasons, it is unlikely that a large amount of dust exists in the outer Solar system to explain the FIR background excess,
therefore it is even unlikely to explain the NIR background excess by the isotropic zodiacal light.

\section{Summary}
This research examined whether the isotropic zodiacal light, if it exists, can explain the observed NIR background excess.
Existence of a spherical IPD shell around the outer Solar system is assumed to produce the isotropic zodiacal light.
From the restriction that thermal emission from the spherical IPD shell must not exceed the observed FIR background excess, the upper limit of the mass of the IPD shell was obtained.
Even if the maximum amount of IPD permissible from the FIR background excess is assumed, 
the isotropic zodiacal light from such IPD shell cannot explain the detected NIR background excess.


\begin{ack}
The author thanks Ko Arimatsu, Shuji Matsuura, Kei Sano, Masakazu Kobayashi, and Masahiro Nagashima for discussions of the IPD in the outer Solar system.
This research is supported by JSPS KAKENHI Grant Number 17K187890.
\end{ack}


\end{document}